\documentclass[a4paper,12pt]{article}
\usepackage{test,tikz,caption,float}

\definecolor{matlabblue}{rgb}{0,0.4470,0.7410}
\definecolor{matlaborange}{rgb}{0.8500,0.3250,0.0980}
\definecolor{matlabyellow}{rgb}{0.9290,0.6940,0.1250}
\definecolor{pantone}{RGB}{1, 33, 105}

\usepackage{hyperref}
\hypersetup{colorlinks,linkcolor=blue,citecolor=blue,urlcolor=pantone}
\usepackage{doi}

\usepackage[style=ext-authoryear-comp,
sorting=nyt, % cite by name, year, title
sortcites=false; % prevent sorting citations
dashed=false, % don't use dash for repeated author
maxcitenames=2, % maximum names to cite in text
maxbibnames=99, % maximum names to list in reference
uniquelist=false,
uniquename=false,
giveninits=true, % first name initials
natbib, % use natbib commands
date=year % year only
]{biblatex}

\AtBeginRefsection{\GenRefcontextData{sorting=ynt}}
\AtEveryCite{\localrefcontext[sorting=ynt]}

\DeclareFieldFormat{pages}{#1} % suppress pp.
\renewbibmacro{in:}{\ifentrytype{article}{}{\printtext{\bibstring{in}\intitlepunct}}} % delete In:

\DeclareFieldFormat[article,inbook,incollection,inproceedings,patent,thesis,unpublished]{titlecase:title}{\MakeSentenceCase*{#1}} % change title to sentence case

\newbibmacro*{atoda:titlelink}[1]{%
  \ifhyperref
    {\iffieldundef{doi}
       {\iffieldundef{eprint}
          {\iffieldundef{url}
             {#1}
             {\href{\thefield{url}}{#1}}}
          {\href{https://arxiv.org/abs/\thefield{eprint}}{#1}}}
       {\href{https://doi.org/\thefield{doi}}{#1}}}
    {#1}}

\DeclareFieldFormat{title}{\usebibmacro{atoda:titlelink}{\mkbibemph{#1}}}
\DeclareFieldFormat
  [article,inbook,incollection,inproceedings,patent,thesis,unpublished]
  {title}{\usebibmacro{atoda:titlelink}{\mkbibquote{#1\isdot}}}

\renewbibmacro*{doi+eprint+url}{}

\newcommand{\aff}{\operatorname{aff}}
\newcommand{\hyp}{\operatorname{hyp}}
\newcommand{\spn}{\operatorname{span}}
\newcommand{\supconv}{\mathbin{\square}}

\usepackage[inline,shortlabels]{enumitem}
\setlist[enumerate,1]{label=(\roman*)}

\usepackage[margin = 1.25in]{geometry}
\usepackage{setspace}
\title{Linearity, Geometry, and Substitution in Aggregate Production}
\author{Christopher P. Chambers\thanks{Department of Economics, Georgetown University. \href{mailto:cc1950@georgetown.edu}{cc1950@georgetown.edu}.} \and Alexis Akira Toda\thanks{Department of Economics, Emory University and Research Institute for Economics and Business Administration, Kobe University. \href{mailto:alexis.akira.toda@emory.edu}{alexis.akira.toda@emory.edu}.}}

\numberwithin{equation}{section}

\begin{document}
\maketitle

\begin{abstract}

We study aggregate production under efficient input allocation across heterogeneous production units. With constant returns to scale, every profit-maximizing allocation generates a cone on which the aggregate production function is linear, whose dimension is the rank of active units' input vectors. Under concavity, the maximal cone at a supporting price is generated by component-technology contact sets. Higher-dimensional cones exist exactly when the pointwise maximum of strictly concave component technologies is nonconcave on the unit simplex. As an application, coexistence of land-using and land-free activities implies infinite aggregate elasticity of substitution at sufficiently high nonland-to-land ratios.

\medskip


\medskip


\end{abstract}

\section{Introduction}

How does an aggregate production technology arise from heterogeneous component
technologies?  Consider $J$ independently scalable production units that use
$N$ common inputs to produce a common, additive output.  Given an aggregate
input vector $x$, one may allocate the inputs across these units to maximize
total output.  The resulting value function $F(x)$ is the \emph{aggregate
production function} studied in this paper.  This definition is appropriate
when inputs are mobile across production units and only total output matters.
It also has a decentralized interpretation: because profits are additive,
maximizing aggregate profit at common prices is equivalent to maximizing the
profit of each unit separately.  For brevity, we refer to the indexed production
units as firms in the general analysis; depending on the application, however,
a unit may represent a firm, plant, sector, or production activity.  The
question is what properties this allocation problem imposes on $F$, even when
the component technologies are heterogeneous and otherwise unrestricted.

One property of particular interest is curvature, which governs the extent to
which one input can replace another.  \citet{HiranoToda2025PNAS}, for example,
study an economy in which land serves as both a factor of production and a store of value.  Their
Land Overvaluation Theorem assumes that the aggregate elasticity of
substitution between land and nonland factors exceeds $1$ when the
nonland-to-land input ratio is high.  This condition concerns the aggregate
technology, although production is carried out by heterogeneous production
units operating within and across sectors.  More generally,
\citet{OberfieldRaval2021} show that aggregate
capital--labor substitution reflects both substitution within plants and the
reallocation of inputs across plants; in U.S. manufacturing, their estimated
industry-level elasticities exceed the corresponding plant-level estimates.
These results motivate a systematic study of how efficient reallocation shapes
the curvature of aggregate production.

Our starting result is that constant returns to scale generate exact linearity on
endogenous regions of the aggregate input space.  Suppose each component
production function $F_j$ is homogeneous of degree one.  Let $x^*$ maximize
aggregate profit at input prices $w$, and let $(x_j^*)_{j=1}^J$ be any
allocation attaining $F(x^*)$.  Theorem \ref{thm:linear_profit} shows that
\begin{equation*}
    F(x)=w\cdot x
    \qquad\text{for every }x\in\cone\set{x_1^*,\dots,x_J^*},
\end{equation*}
where the cone consists of all nonnegative linear combinations of the
input vectors assigned to the production units in the optimal allocation.  This
result does not require concavity.  If the component technologies are also
concave, every strictly
positive aggregate input admits a supporting price vector, and Corollary
\ref{cor:linear_output} gives the same conclusion at any
$\bar{x}\in\R_{++}^N$.

The argument follows from constant returns and zero profit.  Suppose, for
example, that two active firms use input vectors $x_1^*$ and $x_2^*$ at common
input prices $w$.  Scaling firm $j$'s input vector by any $\alpha_j\ge0$
scales its output by the same amount and preserves zero profit.  Hence the
allocation $(\alpha_1x_1^*,\alpha_2x_2^*)$ produces $w\cdot x$ at every
aggregate input
\begin{equation*}
    x=\alpha_1x_1^*+\alpha_2x_2^*.
\end{equation*}
Profit maximization also implies the upper bound $F(x)\le w\cdot x$.
Therefore the scaled allocation is efficient and $F(x)=w\cdot x$.  Changes in
aggregate inputs within the cone can thus be accommodated by changing the
scales of active firms rather than their input proportions.

Linearity along a single ray follows directly from homogeneity.  The dimension
of the cone generated by an allocation is exactly the dimension of the span of
the input vectors in the allocation.  It is therefore full-dimensional if and only
if those vectors span $\R^N$.  Under concavity, Theorem
\ref{thm:maximal_cone} gives an allocation-independent characterization.  At
a supporting price $w$, the maximal cone on which $F(x)=w\cdot x$ is the sum
of the component zero-profit cones.  It is full-dimensional if and only if
the component contact sets contain $N$ affinely independent input shares.  This
criterion distinguishes economically substantive regions from the ray
linearity implied by homogeneity alone.

The geometric characterization supplies a condition for higher-dimensional
linearity cones stated directly in terms of the component technologies.  On the
unit simplex, Theorem \ref{thm:geo} shows that $F$ is the least concave majorant
of their pointwise maximum.  Theorem \ref{thm:M} shows that
nonconcavity of this maximum generates a higher-dimensional linearity cone.  If
the normalized component technologies are strictly concave, the condition is
also necessary; with two
inputs, the resulting cone is full-dimensional.  However, constant returns alone do
not guarantee such a region: for example, if one strictly concave technology dominates all
others, aggregation simply reproduces that technology.  As aggregate inputs
move across contact cones, the active production units or their input vectors may change,
so aggregate curvature can reflect changes in efficient specialization as
well as curvature within individual technologies.

The land application in Section \ref{sec:land} shows how a linearity cone can
extend to an economically relevant boundary of the input space.  With a
nonland input $H$ and land $X$, suppose an efficient allocation operates both a
land-using activity and an activity that uses $H$ but no land.  The two input vectors
generate every aggregate input satisfying $H\ge\kappa X$ for an endogenous
threshold $\kappa$.  The aggregate production function is linear throughout
this region, and its elasticity of substitution is infinite.  Thus the high-input elasticity condition in
\citet{HiranoToda2025PNAS} can arise from reallocation between land-using and
land-free activities even when the land-using technology has finite
elasticity.  The corollary supplies a production-side foundation for that
technological condition; the other assumptions of their land-overvaluation
result concern productivity growth and asset markets.

The hypograph formulation complements these results: the hypograph of
$F|_{\Delta_N}$ is the convex hull of the union of the hypographs of the
component technologies.  Proposition \ref{prop:operate} also gives a sparsity
result.  With $N$ inputs, maximal output at any aggregate input can be attained
using at most $N$ active production units,
although other efficient allocations may use more.

The contribution of the paper is to connect three levels of description that
are usually considered separately: efficient production-unit allocations, exposed
faces of the aggregate technology, and the concavification of the pointwise
maximum of the component technologies.  Theorem \ref{thm:linear_profit} maps any
profit-maximizing allocation into a linearity cone and gives its exact
dimension; Theorem \ref{thm:maximal_cone} identifies the maximal such cone at
a supporting price without selecting an allocation; and Theorems
\ref{thm:geo} and \ref{thm:M} translate the contact geometry
into a condition on the primitive component technologies, with a converse under
strict concavity.  This chain distinguishes the paper's result from the ray
linearity implied by homogeneity alone and yields both the sparsity result and
the primitive condition for high-ratio land--nonland
substitution.

Our efficient-allocation formulation has its roots in activity analysis.
\citet{Koopmans1951} studies production as an efficient combination of
activities, and \citet{Houthakker1955} shows that a continuum of fixed-coefficient
production units with suitably distributed coefficients and capacity
constraints can generate a Cobb--Douglas aggregate production function.
Subsequent work derives other aggregate functional forms or studies
restrictions connecting micro and aggregate technologies
\citep{Levhari1968,Sato1969,Sato1975,Cornwall1973,Vohra1981}.  Activity-based
constructions also appear in macroeconomics and growth theory
\citep{Jones2005,Lagos2006}.  These contributions obtain particular global
aggregate technologies by imposing structure on the form or distribution of
the underlying activities.  We instead impose neither a parametric functional
form nor distributional assumptions across production units and ask which
higher-dimensional linear regions efficient allocation generates, how large
they are, and whether their existence can be inferred from the component
technologies.  For a finite menu of techniques, our geometric characterization
also clarifies the relation to
\citet{Jones2005}: efficient allocation replaces the pointwise envelope on the
unit simplex by its least concave majorant.

Several papers are especially close to ours but impose different technological
restrictions.  \citet{Hildenbrand1981} constructs a short-run industry
technology as the Minkowski sum of capacity-bounded, fixed-coefficient
activities.  The bounds preclude constant returns for this industry frontier;
we instead assume constant returns at the production-unit level and study linearity cones
without capacity bounds.
\citet{Cordoba2013} obtains linear segments in a two-technology model of
structural change: movement along a segment reallocates labor from a backward
sector to an advanced sector and prevents diminishing marginal returns to
capital during the transition.  Our results give general conditions for such
segments and their higher-dimensional analogues.  \citet{BlundellGreenJin2022}
study firms choosing between centralized and decentralized organizational
technologies that use skilled and unskilled labor.  When both technologies
operate, their aggregate production function has a linear region associated
with a diversification cone.  Our results isolate the general contact-cone
geometry behind this mechanism and characterize it for an arbitrary finite
number of inputs and technologies.  \citet{CordobaRipoll2009}
solve an efficient-allocation problem for agriculture and nonagriculture with
Cobb--Douglas sectoral technologies and land, capital, and labor.  Their
diversification component measures the gain from operating both sectors and is
used for development accounting.  Our land application instead asks
when land-using and land-free activities produce a high-input-ratio linearity
cone and infinite aggregate elasticity of substitution.  It is also related
to \citet{HansenPrescott2002}, whose growth model combines a Malthus technology
using land, labor, and reproducible capital with a Solow technology that does
not use land.  We abstract from their dynamic transition and identify a
primitive condition under which coexistence of land-using and land-free
activities generates a high-ratio linearity cone.

Our question must also be distinguished from the classical exact-aggregation
problem studied by \citet{May1946,Fisher1969} and reviewed by
\citet{FelipeFisher2003}.  That literature asks whether a technology described
in terms of many heterogeneous inputs or outputs can be represented using a
smaller number of quantity indices---for example, whether firm-specific
capital goods can be summarized by a price-independent capital aggregate.
Even when elementary inputs are allocated efficiently, such a dimensional
reduction requires stringent separability conditions.  We start after this
index-number problem: each coordinate of $x$ denotes a common input measured
in the same units across production units, and the units produce a common, additive
output.  Supremal convolution then defines maximum total output for each vector
of these common inputs.  It does not establish that heterogeneous capital,
labor, or output goods admit scalar aggregates; if a coordinate of $x$ is
itself such an index, the classical aggregation restrictions continue to
apply.  \citet{BaqaeeFarhi2019} take a broader general-equilibrium approach to
aggregate production with heterogeneous producers, factors, and input-output
linkages.  Our partial-equilibrium, common-output setting isolates the role of
efficient input reallocation.

The same allocation problem yields the GDP, or revenue, function when sectors
produce distinct goods and output prices are fixed.  Multiplying each sector's
production function by its output price converts value maximization into the
common-output formulation used here.  The resulting linearity cones are
closely related to diversification cones in trade theory.  Within a
full-dimensional diversification cone, zero-profit conditions determine
factor prices, while changes in factor endowments are accommodated by changes
in sectoral output---the basis of factor-price insensitivity and the
Rybczynski theorem
\citep{Samuelson1948,Samuelson1953,McKenzie1955,Rybczynski1955,Leamer1995}.
Our results express the same geometry as a property of the aggregate
production function.  This cone-wise constancy differs from the classical
non-substitution theorem, which obtains global price independence under
stronger assumptions \citep{Samuelson1951}.

\section{Preliminaries}\label{sec:prelim}

This section defines the aggregate production function and records the basic properties used below.

Suppose that there are $N$ inputs indexed by $n=1,\dots,N$ and one output. For the purposes of this paper, a \emph{production function} is any upper semicontinuous (usc) function $F:\R_+^N\to \R_+$. We interpret $F(x)$ as the frontier output attainable when the input vector $x$ is assigned to a production unit.

Suppose that there are $J$ production units, which we call firms, indexed by $j=1,\dots,J$, with unit $j$ having production function $F_j$. We define the \emph{aggregate production function} as the maximum output that a central planner can achieve by optimally assigning an input vector $x_j\in\R_+^N$ to each firm $j$, subject to the exact resource constraint $\sum_{j=1}^Jx_j=x$.\footnote{We permit free disposal of output but not that of input, which here is equivalent to coordinatewise monotonicity. If every $F_j$ is monotone, the equality constraint may be replaced by $\sum_jx_j\le x$, because any unused input can be assigned without reducing output.} Formally, we define $F:\R_+^N\to \R_+$ by
\begin{equation}
    F(x)\coloneqq \max\set*{\sum_{j=1}^JF_j(x_j): x_j\in\R_+^N\text{ for every }j, \sum_{j=1}^Jx_j=x}. \label{eq:agg_prod_func}
\end{equation}
For every $x$, the feasible set in \eqref{eq:agg_prod_func} is nonempty and compact, and the objective function is usc. Hence the maximum is attained. Because each $F_j$ is nonnegative, $F$ maps $\R_+^N$ into $\R_+$. Furthermore, the upper-semicontinuity part of Berge's maximum theorem (\citealp[Lemma 17.30]{AliprantisBorder2006}) implies that $F$ is usc. Hence $F$ is a production function under our definition.

The operation in \eqref{eq:agg_prod_func} is known as the \emph{supremal convolution} \citep[p.~34]{Rockafellar1970}. Throughout the rest of the paper, let $(F_j)_{j=1}^J$ be production functions and write $F\coloneqq F_1\supconv \dotsb \supconv F_J$ for the aggregate production function defined in \eqref{eq:agg_prod_func}. We also use the Minkowski sum of sets.  For subsets $A_1,\dots,A_J$ of a vector space, define
\begin{equation*}
    A_1+\dotsb+A_J
    \coloneqq\set{a_1+\dotsb+a_J:a_j\in A_j\text{ for every }j}.
\end{equation*}
The following standard result shows that individual and aggregate profit maximization are equivalent at common input and output prices.

\begin{prop}[Aggregate and firm profit maximization]\label{prop:agg}
Let $w\in \R_+^N$ be an input price vector and normalize the output price to $1$. Then
\begin{equation*}
    x^*\in \argmax_{x\in\R_+^N}(F(x)-w\cdot x)
\end{equation*}
if and only if there exist $(x_j^*)_{j=1}^J$ with $x^*=\sum_{j=1}^J x_j^*$ such that, for all $j$,
\begin{equation*}
    x_j^*\in \argmax_{x\in\R_+^N}(F_j(x)-w\cdot x).
\end{equation*}
\end{prop}

\begin{proof}
Let $Y_j\coloneqq \set*{(-x,u)\in \R^{N+1}:x\in\R_+^N,u\le F_j(x)}$ be the production set of firm $j$ and define the aggregate production set $Y$ analogously. It follows from \eqref{eq:agg_prod_func} that $Y=Y_1+\dotsb+Y_J$, so the result is immediate from \citet[Proposition 13.1]{Kreps2013}.
\end{proof}

Proposition \ref{prop:agg} implies that, when only aggregate quantities matter and inputs are mobile across firms, the industry can be represented by a single aggregate firm. The same aggregation operator yields the GDP, or revenue, function at fixed output prices. Suppose that sector $j$ produces a distinct good according to $f_j:\R_+^N\to\R_+$, and let $p_j>0$ denote its output price. Given an aggregate factor endowment $v\in\R_+^N$, the maximum value of production is
\begin{equation*}
    G(p,v) \coloneqq \max\set*{\sum_{j=1}^J p_j f_j(v_j): v_j\in\R_+^N\text{ for every }j, \sum_{j=1}^J v_j=v}.
\end{equation*}
Letting $F_j\coloneqq p_j f_j$, we obtain $G(p,v)=(F_1\supconv \dotsb \supconv F_J)(v)$. Conversely, our common-output aggregate production function is the special case in which all output prices equal $1$. Thus, for fixed $p$, the results below apply directly to $G(p,\cdot)$ whenever the sectoral technologies satisfy the corresponding assumptions. In the trade interpretation, the supporting input prices are competitive factor prices, and the resulting linearity cones correspond to diversification cones.

We introduce more structure to production functions. A production function $F$ exhibits constant returns to scale, or equivalently is homogeneous of degree one (or simply homogeneous), if $F(\lambda x)=\lambda F(x)$ for all $\lambda \ge 0$. Concavity and homogeneity are both preserved by aggregation: concavity follows from \citet[Theorem 5.4]{Rockafellar1970}, while homogeneity follows directly from \eqref{eq:agg_prod_func}. Concavity and homogeneity imply that $F$ is superadditive. Indeed, for any $x,y\in\R_+^N$,
\begin{equation}
    F(x+y)=2F\left(\frac{x+y}{2}\right) \ge F(x)+F(y). \label{eq:superadditive}
\end{equation}

For $x\in \R^N$, let $\norm{x}\coloneqq \sum_{n=1}^N \abs{x_n}$ denote the standard $\ell^1$ norm and 
\begin{equation*}
    \Delta_N \coloneqq \set*{x\in \R_+^N: \norm{x}=1}
\end{equation*}
the unit simplex. Since a homogeneous function $F$ satisfies $F(x)=\norm{x}F(x/\norm{x})$ for $x\neq 0$, it is uniquely determined by its restriction $F|_{\Delta_N}$ to $\Delta_N$. We often use the notation $z=x/\norm{x}$ for the normalized input vector.

When $N=2$, we parameterize the simplex by
\begin{equation*}
    z(t)\coloneqq(t,1-t),
    \qquad t\in[0,1].
\end{equation*}

The following lemma shows that the calculation of the aggregate production function of homogeneous production functions reduces to that on the simplex.

\begin{lem}[Simplex representation]\label{lem:agg_prod_homog}
If each $F_j$ is homogeneous, then for any $z\in \Delta_N$, the maximum is taken over weights $\alpha\in\Delta_J$ and normalized input vectors $z_j\in\Delta_N$, one for each firm, satisfying $\sum_{j=1}^J\alpha_jz_j=z$.  In particular,
\begin{equation}
    F(z)=\max\set*{\sum_{j=1}^J \alpha_j F_j(z_j): \alpha\in \Delta_J,\ z_j\in\Delta_N\text{ for every }j,\ \sum_{j=1}^J \alpha_jz_j=z}. \label{eq:agg_prod_homog}
\end{equation}
\end{lem}

\section{Linearity and geometry of \texorpdfstring{$F$}{F}}\label{sec:linear}

In this section we present our main results. We first show that a profit-maximizing allocation generates a cone on which the aggregate production function is linear and determine the dimension of that cone. Under concavity, the same conclusion applies at every strictly positive aggregate input.  We then characterize the maximal cone associated with a supporting price and give necessary and sufficient conditions for it to be full-dimensional. Finally, we characterize the aggregate technology on the unit simplex and determine when heterogeneity generates a genuinely higher-dimensional linearity cone.

We use the following terminology.  A convex cone $C\subset\R_+^N$ is a \emph{linearity cone} of $F$ if there exists $a\in\R^N$ such that $F(x)=a\cdot x$ for every $x\in C$.  Its dimension is $\dim C\coloneqq\dim\spn C$. The cone is
\emph{higher-dimensional} if $\dim C\ge2$ and
\emph{full-dimensional} if $\dim C=N$, or equivalently if it has nonempty
interior in $\R^N$.

\subsection{Allocation-generated cones}

For a subset $A$ of a Euclidean space, let $\conv A$ denote its convex hull.
For vectors $v_1,\dots,v_K\in\R^N$, define their conical hull by
\begin{equation*}
    \cone\set{v_1,\dots,v_K}\coloneqq \set*{\sum_{k=1}^K \alpha_k v_k : \alpha_k\ge 0}.
\end{equation*}

\begin{thm}[Linearity under profit maximization]\label{thm:linear_profit}
Suppose that each $F_j$ is homogeneous. Let $w\in\R_+^N$ be an input price vector, and suppose that
$x^*$ maximizes aggregate profit at prices $w$. Let $(x_j^*)_{j=1}^J$ be any allocation attaining $F(x^*)$, so that $\sum_{j=1}^Jx_j^*=x^*$ and $\sum_{j=1}^JF_j(x_j^*)=F(x^*)$. Define the cone $C\coloneqq \cone\set{x_1^*,\dots,x_J^*}$. Then $F(x)=w\cdot x$ for every $x\in C$, and
\begin{equation*}
    \dim C=\dim\spn \set{x_1^*,\dots,x_J^*}.
\end{equation*}
Consequently, $C$ is full-dimensional if and only if the input vectors in the allocation span $\R^N$.  In that case, the coefficient vector $w$ is unique.
\end{thm}

\begin{proof}%[Proof of Theorem \ref{thm:linear_profit}]
Because the individual production functions $(F_j)_{j=1}^J$ are homogeneous, so is $F$.  Hence the profit objective $F(x)-w\cdot x$ is also homogeneous.

Since $F(0)=0$, the maximal profit is nonnegative. It cannot be strictly positive: if $F(x^*)-w\cdot x^*>0$, then homogeneity would imply
\begin{equation*}
    F(2x^*)-w\cdot(2x^*)
    =2[F(x^*)-w\cdot x^*]
    >F(x^*)-w\cdot x^*,
\end{equation*}
contradicting the optimality of $x^*$. Therefore $F(x^*)-w\cdot x^*=0$. Because $x^*$ maximizes profit, it follows that $F(x)-w\cdot x\le 0$ for all $x\in \R_+^N$, or equivalently,
\begin{equation}
	(\forall x\in \R_+^N)\quad F(x)\le w\cdot x. \label{eq:F_ub}
\end{equation}

Fix $j$ and evaluate the aggregate production function at the aggregate input $x_j^*$, rather than at $x^*$.  Allocating the entire vector $x_j^*$ to firm $j$ and zero inputs to every other firm exhausts this aggregate input, so it is feasible in \eqref{eq:agg_prod_func}. Thus
\begin{equation*}
    F_j(x_j^*)\le F(x_j^*)\le w\cdot x_j^*,
\end{equation*}
where the second inequality follows from \eqref{eq:F_ub}. Summing these inequalities gives
\begin{equation*}
    \sum_{j=1}^J F_j(x_j^*) \le \sum_{j=1}^J w\cdot x_j^*.
\end{equation*}
Both sides are in fact equal, because
\begin{equation*}
    \sum_{j=1}^J F_j(x_j^*)=F(x^*)=w\cdot x^*=\sum_{j=1}^J w\cdot x_j^*.
\end{equation*}
It follows that
\begin{equation}
(\forall j)\quad F_j(x_j^*)=w\cdot x_j^*.
\label{eq:firm-zero-profit}
\end{equation}

Now take any $x\in C$. By definition, there exist coefficients $\alpha_j\ge 0$ such that $x=\sum_{j=1}^J \alpha_jx_j^*$. The allocation in which firm $j$ receives $\alpha_jx_j^*$ is feasible for the aggregate input vector $x$. Therefore, using the homogeneity of each $F_j$ and \eqref{eq:firm-zero-profit},
\begin{equation}
    F(x)\ge \sum_{j=1}^J F_j(\alpha_jx_j^*)=\sum_{j=1}^J \alpha_jF_j(x_j^*)=\sum_{j=1}^J \alpha_j w\cdot x_j^*=w\cdot x. \label{eq:F_lb}
\end{equation}
Combining \eqref{eq:F_ub} and \eqref{eq:F_lb}, we obtain $F(x)=w\cdot x$ for all $x\in C$.

Finally, since $C$ is generated by $(x_j^*)_{j=1}^J$, $\dim C=\dim\spn \set{x_1^*,\dots,x_J^*}$. Thus $C$ is full-dimensional if and only if these vectors span $\R^N$. If another vector $a\in\R^N$ satisfies $F(x)=a\cdot x$ for all $x\in C$, then $(a-w)\cdot x=0$ on $C$, implying $a=w$ because $C$ is full-dimensional.
\end{proof}

Theorem \ref{thm:linear_profit} applies to an aggregate input that maximizes profit for some input price vector $w$. Without concavity, a given aggregate input $\bar{x}\in\R_+^N$ need not have this property, as the following example shows.

\begin{exmp}[Failure of support without concavity]\label{ex:nonconcave}
Suppose there are $N=2$ inputs and $J=2$ firms with homogeneous production functions $F_1,F_2$, and write $\bar{x}=z(\bar t)\in\Delta_2$.  The functions $F_j$ are uniquely determined by the curves $t\mapsto F_j(z(t))$ on $[0,1]$. Suppose these curves are given by the graphs in the left panel of Figure \ref{fig:prod_simplex}. The right panel repeats these curves with thinner strokes and superimposes the aggregate production function as a thick black curve.

One can verify from Lemma \ref{lem:agg_prod_homog} that the black polygonal curve in the right panel of Figure \ref{fig:prod_simplex} is $F=F_1\supconv F_2$ on $\Delta_2$.  Take $\bar t=7/10$, as indicated by the vertical dashed line. The dotted horizontal segment connects the firm-1 activity at $t=1/5$ to the firm-2 activity at $t=9/10$, which produce the same output.  Every point on this segment is attainable by mixing and scaling these two activities.  The segment meets the aggregate graph at $\bar t$, so the corresponding allocation attains $F(\bar{x})$, but this point lies strictly below the aggregate graph near $t=2/5$.  Consequently, the segment cannot be extended to an affine upper support of $F|_{\Delta_2}$.  Equivalently, the aggregate graph at $\bar t$ lies strictly below its concave envelope. Hence $\bar{x}$ cannot maximize $F(x)-w\cdot x$ for any input price vector $w$, and Theorem \ref{thm:linear_profit} cannot be applied.

\begin{figure}[!htb]
\centering

\begin{tikzpicture}[scale = 0.6]

\draw (0,0) -- (10,0);
\draw (5,0) node[below] {$t$};

% F1
\draw[thick,color=matlabblue] (0,0) -- (2,5) -- (4,0) -- (10,0);
\draw[color=matlabblue] (2,5) node[above] {$F_1$};
% F2
\draw[thick,dashed,color=matlaborange] (0,0) -- (4,6) -- (8,0) -- (9,5) -- (10,0);
\draw[color=matlaborange] (4,6) node[above] {$F_2$};

\end{tikzpicture}
\quad
\begin{tikzpicture}[scale = 0.6]

\draw (0,0) -- (10,0);
\draw (5,0) node[below] {$t$};

% F1
\draw[color=matlabblue] (0,0) -- (2,5) -- (4,0) -- (10,0);
\draw[color=matlabblue] (2,5) node[above] {$F_1$};
\draw[densely dotted,color=matlabblue] (2,5) -- (9,5);
% F2
\draw[dashed,color=matlaborange] (0,0) -- (4,6) -- (8,0) -- (9,5) -- (10,0);
\draw[color=matlaborange] (4,6) node[above] {$F_2$};
% F
\draw[thick] (0,0) -- (2,5) -- (4,6) -- (5,5) -- (9,5) -- (10,0);
\draw (7,5) node[above] {$F=F_1\supconv F_2$};
% \bar t
\draw[dashed] (7,0) -- (7,5);
\draw (7,0) node[below] {$\bar t$};

\end{tikzpicture}

\caption{Production functions on the simplex $\Delta_2$.}\label{fig:prod_simplex}
\caption*{\footnotesize Note: In this and other figures, the horizontal axis is $t\in[0,1]$, representing the simplex point $z(t)=(t,1-t)$, the vertical axis shows the relevant function value, and component technologies are distinguished by colors and line styles.}
\end{figure}
\end{exmp}

However, under concavity, every strictly positive aggregate input admits a supporting price vector, so Theorem \ref{thm:linear_profit} applies at an arbitrary $\bar{x}\in\R_{++}^N$. To state this result, recall that for a concave function $F$, a vector $w$ satisfying
\begin{equation}
    F(x)-F(\bar{x})\le w\cdot(x-\bar{x})\label{eq:supergradient}
\end{equation}
for all $x$ is called a supergradient at $\bar{x}$. The set of supergradients at $\bar{x}$ is denoted by $\partial F(\bar{x})$. Applying Theorem 23.4 of \citet{Rockafellar1970} to the convex function $-F$, we have $\partial F(\bar{x})\neq \emptyset$ whenever $\bar{x}$ is in the interior of the domain of $F$. The following corollary identifies $w\in \partial F(\bar{x})$ as a coefficient vector for $F$ on the cone generated by an optimal allocation.

\begin{cor}[Linearity under output maximization]\label{cor:linear_output}
Suppose each $F_j$ is homogeneous and concave. Let $\bar{x}\in \R_{++}^N$ and let $(x_j^*)_{j=1}^J$ be any allocation attaining $F(\bar{x})$. Then, for any $w\in \partial F(\bar{x})$, the conclusion of Theorem \ref{thm:linear_profit} holds with $x^*=\bar{x}$.
\end{cor}

\begin{proof}
Since each $F_j$ is concave, so is $F$. Since $\bar{x}\in \R_{++}^N$ is in the interior of the domain of $F$, we have $\partial F(\bar{x})\neq \emptyset$. Take any $w\in \partial F(\bar{x})$. Setting $x=0$ and $x=2\bar{x}$ in \eqref{eq:supergradient} and using homogeneity gives $F(\bar{x})=w\cdot\bar{x}$. Hence $F(x)\le w\cdot x$ for every $x\in\R_+^N$ by \eqref{eq:supergradient}. In particular, $w_n\ge F(e_n)\ge0$ for every $n$ (where $e_n$ is the $n$-th unit vector), so $w$ can
be interpreted as an input price vector. Moreover, $\bar{x}$ maximizes $F(x)-w\cdot x$. The conclusion now follows from Theorem \ref{thm:linear_profit}.
\end{proof}

\subsection{Maximal contact cones}

The cones in Theorem \ref{thm:linear_profit} and Corollary \ref{cor:linear_output} depend on a particular optimal allocation.  We next collect all input vectors that earn zero profit at the same supporting price. Suppose that $w\in\R_+^N$ satisfies $F(x)\le w\cdot x$ for every $x\in\R_+^N$, and define the component and aggregate \emph{contact sets} (the sets of profit maximizers)
\begin{equation*}
    Z_j(w)\coloneqq\set*{x\in\R_+^N:F_j(x)=w\cdot x},
    \qquad
    Z(w)\coloneqq\set*{x\in\R_+^N:F(x)=w\cdot x}.
\end{equation*}
Define the normalized contact sets by
\begin{equation*}
    \widehat{Z}_j(w)\coloneqq Z_j(w)\cap\Delta_N,
    \qquad
    \widehat{Z}(w)\coloneqq Z(w)\cap\Delta_N.
\end{equation*}
The following theorem shows that the aggregate contact set is the maximal set on which $F$ has the particular
linear representation $w\cdot x$.

\begin{thm}[Maximal contact cone]\label{thm:maximal_cone}
Suppose that each $F_j$ is homogeneous and that $F(x)\le w\cdot x$ for every
$x\in\R_+^N$. Then the following statements hold.
\begin{enumerate}
    \item\label{item:maximal_cone1} The aggregate contact set satisfies $Z(w)=Z_1(w)+\dotsb+Z_J(w)$.

    \item\label{item:maximal_cone2} If each $F_j$ is concave, then the sets $Z_j(w)$ and $Z(w)$ are
    closed convex cones, and
    \begin{equation}
        \widehat{Z}(w)
        =\conv\left(\bigcup_{j=1}^J\widehat{Z}_j(w)\right).
        \label{eq:Zhatw}
    \end{equation}

    \item\label{item:maximal_cone3} If each $F_j$ is concave and $Z(w)\neq\set{0}$, then
    \begin{equation*}
        \dim Z(w)=1+\dim\aff \widehat{Z}(w).
    \end{equation*}
    Consequently, $Z(w)$ is full-dimensional if and only if
    $\bigcup_{j=1}^J\widehat{Z}_j(w)$ contains $N$ affinely independent points.
\end{enumerate}
\end{thm}

The theorem can be read in terms of profit shortfalls.  Assigning all inputs
to firm $j$ shows that $F_j(x)\le F(x)\le w\cdot x$, so
$w\cdot x-F_j(x)$ is nonnegative.  Moreover, because
$w\cdot x=\sum_jw\cdot x_j$ for every feasible allocation,
\begin{equation*}
    w\cdot x-F(x)
    =\min_{\substack{x_j\in\R_+^N\\ \sum_{j=1}^Jx_j=x}}
    \sum_{j=1}^J\bigl[w\cdot x_j-F_j(x_j)\bigr].
\end{equation*}
The aggregate shortfall is therefore zero if and only if an allocation can be chosen so that every term $w\cdot x_j-F_j(x_j)$ is zero.  Equivalently, $x\in Z(w)$ if and only if $x=\sum_jx_j$ with $x_j\in Z_j(w)$, which is precisely the Minkowski-sum identity in Theorem \ref{thm:maximal_cone}\ref{item:maximal_cone1}.

The simplex formula \eqref{eq:Zhatw} has an equally direct interpretation.  For a normalized aggregate input $x\in \widehat{Z}(w)$, write each nonzero $x_j$ as $x_j=\alpha_jz_j$, where $z_j\in \widehat{Z}_j(w)$ and $\alpha_j=\norm{x_j}$.  The coefficients satisfy $\sum_j\alpha_j=1$, so the aggregate input share $x$ is a convex combination of the component contact shares $z_j$. Thus changes in aggregate inputs within $Z(w)$ can be absorbed by changing the scales and mixture of zero-profit activities while the same vector $w$ continues to support aggregate production.  A full-dimensional contact cone means that this adjustment is possible over an open set of aggregate input vectors, rather than only along a ray or a lower-dimensional set.

\begin{exmp}[Two distinct contact rays]\label{ex:two_contact_rays}
Suppose $N=J=2$ and, at a supporting price $w$, each component technology has a unique normalized contact point $z_j\in\Delta_2$, with $z_1\ne z_2$.  Then $\widehat{Z}_j(w)=\set{z_j}$ and homogeneity gives $Z_j(w)=\cone\set{z_j}$. Theorem \ref{thm:maximal_cone} therefore implies
\begin{equation*}
    Z(w)=\cone\set{z_1,z_2},
    \qquad
    \widehat{Z}(w)=\conv\set{z_1,z_2}.
\end{equation*}
Write $z_j=z(\theta_j)$ with $\theta_1<\theta_2$.  For an aggregate input $x=(u,v)\ne0$, aggregate production equals $w\cdot x$ whenever
\begin{equation*}
    \theta_1\le \frac{u}{u+v}\le\theta_2.
\end{equation*}
An input-share change within this interval is accommodated by changing the relative scales of the two activities.  Outside it, at least one activity must depart from its contact ray, and aggregate output lies strictly below $w\cdot x$. Example \ref{ex:cobb_douglas} and Figure \ref{fig:CD} give a concrete Cobb--Douglas instance of this construction.
\end{exmp}

The maximal contact cone in Theorem \ref{thm:maximal_cone} can be larger than the cone generated by a particular optimal allocation.  If $F_j|_{\Delta_N}$ is strictly concave, then $\widehat{Z}_j(w)$ contains at most one point: a supporting affine function cannot touch a strictly concave function at two distinct points.  More generally, whenever each active component technology has a unique contact point, a full-dimensional cone requires at least $N$ active component technologies with affinely independent input shares. Thus, under this uniqueness condition, $J<N$ rules out a full-dimensional allocation-generated cone.

\subsection{Geometry and existence}

We next give a geometric characterization of the aggregate production function. For any function $f:D\to\R$, define its hypograph by
\begin{equation*}
    \hyp f\coloneqq\set{(x,y):x\in D,\ y\le f(x)}.
\end{equation*}

\begin{thm}[Geometric characterization of $F$]\label{thm:geo}
If each $F_j$ is homogeneous and concave, then
\begin{equation*}
    \hyp F|_{\Delta_N} = \conv \left(\bigcup_{j=1}^J \hyp F_j|_{\Delta_N}\right).
\end{equation*}
Equivalently, $F|_{\Delta_N}$ is the least concave majorant of the functions
$(F_j|_{\Delta_N})_{j=1}^J$.
\end{thm}

Lemma \ref{lem:agg_prod_homog} supplies the geometry.  Every efficient allocation on $\Delta_N$ represents an aggregate hypograph point as a convex combination of points in the component hypographs.  Conversely, grouping any such convex combination by component technology and using concavity and homogeneity produces a feasible allocation with at least the proposed output.

To express Theorem \ref{thm:geo} directly in terms of the component
technologies, define their pointwise maximum on the simplex by
\begin{equation}
    M(z)\coloneqq\max_{1\le j\le J}F_j(z),
    \qquad z\in\Delta_N.
    \label{eq:Mz}
\end{equation}
A function majorizes every $F_j|_{\Delta_N}$ if and only if it majorizes $M$.
Hence Theorem \ref{thm:geo} identifies $F|_{\Delta_N}$ as the least concave
majorant of $M$. This observation yields the following existence criterion for
higher-dimensional linearity cones.

\begin{thm}[Existence of higher-dimensional linearity cones]
\label{thm:M}
Suppose that each $F_j$ is homogeneous and concave and let $M$ be as in \eqref{eq:Mz}. Then:
\begin{enumerate}
\item\label{item:M1} If $M$ is not concave, then $F$ has a higher-dimensional linearity cone.
\item\label{item:M2} If, in addition, each $F_j|_{\Delta_N}$ is strictly concave, $F$ has a higher-dimensional linearity cone if and only if $M$ is not concave.
\end{enumerate}
\end{thm}

Constant returns alone therefore do not ensure a higher-dimensional linearity cone. A particularly transparent special case is one in which a single concave technology dominates all the others.

\begin{exmp}[Dominant technology]\label{ex:dominant}
Suppose that, for some $k$, the function $F_k$ is homogeneous and concave and $F_j(x)\le F_k(x)$ for all $x\in\R_+^N$ and $j$. Hence every feasible allocation $(x_j)_{j=1}^J$ satisfies
\begin{equation*}
    \sum_{j=1}^JF_j(x_j)
    \le\sum_{j=1}^JF_k(x_j)
    \le F_k\left(\sum_{j=1}^Jx_j\right),
\end{equation*}
where the last inequality uses superadditivity \eqref{eq:superadditive}. Assigning all inputs to production unit $k$ attains the last expression, so $F=F_k$.
If $F_k|_{\Delta_N}$ is strictly concave, its graph contains no nondegenerate
line segment, and the aggregate production function therefore has no
higher-dimensional linearity cone.
\end{exmp}

\begin{exmp}[Cobb--Douglas technologies]\label{ex:cobb_douglas}
A parametric instance of Example \ref{ex:two_contact_rays}, and a contrast to Example \ref{ex:dominant}, arises when there are $N=2$ inputs and $J=2$ production units with Cobb--Douglas production functions $F_j(u,v)=u^{\theta_j}v^{1-\theta_j}$. Figure \ref{fig:CD} shows the curves $t\mapsto F_j(z(t))$ when $\theta_1=1/3$ and $\theta_2=2/3$. Theorem \ref{thm:geo} implies that $F|_{\Delta_2}$ is the least concave majorant of $F_1|_{\Delta_2}$ and $F_2|_{\Delta_2}$.  Their pointwise maximum is not concave: the two functions cross at $t=1/2$, where the slope of the maximum jumps upward from $-1/3$ to $1/3$.  Theorem \ref{thm:M} therefore guarantees a full-dimensional linearity cone.

To identify the maximal cone, define
\begin{equation*}
    c\coloneqq\frac{2^{2/3}}{3},
    \qquad
    w\coloneqq(c,c),
    \qquad
    z_1\coloneqq z(\theta_1)=\left(\frac13,\frac23\right),
    \quad
    z_2\coloneqq z(\theta_2)=\left(\frac23,\frac13\right).
\end{equation*}
Both technologies satisfy $F_j(u,v)\le c(u+v)=w\cdot(u,v)$, with equality exactly on the corresponding contact ray. Hence
\begin{equation*}
    Z_1(w)=\cone\set{z_1},
    \qquad
    Z_2(w)=\cone\set{z_2},
\end{equation*}
and Theorem \ref{thm:maximal_cone} gives
\begin{equation*}
    Z(w)=\cone\set{z_1,z_2},
    \qquad
    \widehat{Z}(w)=\conv\set{z_1,z_2}.
\end{equation*}
The right panel of Figure \ref{fig:CD} shows how these contact points determine the aggregate curve.  The two black dots mark $z_1$ and $z_2$, the points at which the two component technologies meet the common supporting line $w\cdot x=c$ on $\Delta_2$.  The thick black curve coincides with $F_1$ for $t\le\theta_1$, follows the horizontal segment $\conv\set{z_1,z_2}$ for $\theta_1\le t\le\theta_2$, and coincides with $F_2$ for $t\ge\theta_2$.  Thus the horizontal segment is exactly the normalized aggregate contact set $\widehat{Z}(w)$.  Homogeneous extension turns it into the two-dimensional linearity cone on which $F(x)=w\cdot x$.

\begin{figure}[!htb]
\centering

\begin{tikzpicture}[scale = 0.6]

\draw (0,0) -- (10,0);
\draw (5,0) node[below] {$t$};

% F1
% Cubic reparameterization clusters samples near both endpoints.
\draw[thick,domain = 0:1,samples = 96,smooth,color=matlabblue,variable=\t]
plot ({10*\t^3/(\t^3+(1-\t)^3)},{10*\t*(1-\t)^2/(\t^3+(1-\t)^3)});
\draw[color=matlabblue] (8,3.175) node[below left] {$F_1$};

% F2
\draw[thick,dashed,domain = 0:1,samples = 96,smooth,color=matlaborange,variable=\t]
plot ({10*\t^3/(\t^3+(1-\t)^3)},{10*\t^2*(1-\t)/(\t^3+(1-\t)^3)});
\draw[color=matlaborange] (2,3.175) node[below right] {$F_2$};

\end{tikzpicture}
\quad
\begin{tikzpicture}[scale = 0.6]

\draw (0,0) -- (10,0);
\draw (5,0) node[below] {$t$};

% F1
\draw[domain = 0:1,samples = 96,smooth,color=matlabblue,variable=\t]
plot ({10*\t^3/(\t^3+(1-\t)^3)},{10*\t*(1-\t)^2/(\t^3+(1-\t)^3)});
\draw[color=matlabblue] (8,3.175) node[below left] {$F_1$};

% F2
\draw[dashed,domain = 0:1,samples = 96,smooth,color=matlaborange,variable=\t]
plot ({10*\t^3/(\t^3+(1-\t)^3)},{10*\t^2*(1-\t)/(\t^3+(1-\t)^3)});
\draw[color=matlaborange] (2,3.175) node[below right] {$F_2$};

% F
\draw[thick,domain = 0:0.442493334024,samples = 48,smooth,variable=\t]
plot ({10*\t^3/(\t^3+(1-\t)^3)},{10*\t*(1-\t)^2/(\t^3+(1-\t)^3)});
\draw[line width=1pt] ({10/3},{10*2^(2/3)/3}) -- ({20/3},{10*2^(2/3)/3});
\draw[thick,domain = 0.557506665976:1,samples = 48,smooth,variable=\t]
plot ({10*\t^3/(\t^3+(1-\t)^3)},{10*\t^2*(1-\t)/(\t^3+(1-\t)^3)});
\fill ({10/3},{10*2^(2/3)/3}) circle (2pt);
\fill ({20/3},{10*2^(2/3)/3}) circle (2pt);
\draw (5,5.291) node[above] {$F=F_1 \supconv F_2$};

\end{tikzpicture}

\caption{Aggregate production function for two Cobb--Douglas functions.}\label{fig:CD}
\caption*{\footnotesize Note: The black dots in the right panel are the
normalized component contact points $z_1$ and $z_2$.  The horizontal segment between
them is $\widehat{Z}(w)$.}
\end{figure}
\end{exmp}

\begin{exmp}[Leontief technologies]\label{ex:leontief}
Consider three production units with Leontief technologies:
\begin{equation*}
    F_1(u,v)=\min\set{2u,v/2}, \quad
    F_2(u,v)=\min\set{u,v}, \quad
    F_3(u,v)=\min\set{u/2,2v}.
\end{equation*}
On $\Delta_2$, these functions attain their peaks at $t=1/5$, $t=1/2$,
and $t=4/5$, respectively, as shown in the left panel of Figure
\ref{fig:Leontief}.  The right panel repeats the three component curves and adds
$F=F_1\supconv F_2\supconv F_3$ in thick black.  By Theorem \ref{thm:geo},
the black polygonal curve is their least concave majorant: it joins the three
component peaks, together with the common zero-output endpoints at $t=0$ and
$t=1$.

\begin{figure}[!htb]
\centering

\begin{tikzpicture}[scale = 0.6]

\draw (0,0) -- (10,0);
\draw (5,0) node[below] {$t$};

% F1
\draw[thick,color=matlabblue] (0,0) -- (2,4) -- (10,0);
\draw[color=matlabblue] (2,4) node[above left] {$F_1$};

% F2
\draw[thick,dashed,color=matlaborange] (0,0) -- (5,5) -- (10,0);
\draw[color=matlaborange] (5,5) node[below] {$F_2$};

% F3
\draw[thick,dash pattern=on 5pt off 2pt on 1pt off 2pt,color=matlabyellow] (0,0) -- (8,4) -- (10,0);
\draw[color=matlabyellow] (8,4) node[above right] {$F_3$};

\end{tikzpicture}
\quad
\begin{tikzpicture}[scale = 0.6]

\draw (0,0) -- (10,0);
\draw (5,0) node[below] {$t$};

% F1
\draw[color=matlabblue] (0,0) -- (2,4) -- (10,0);
\draw[color=matlabblue] (2,4) node[above left] {$F_1$};

% F2
\draw[dashed,color=matlaborange] (0,0) -- (5,5) -- (10,0);
\draw[color=matlaborange] (5,5) node[below] {$F_2$};

% F3
\draw[dash pattern=on 5pt off 2pt on 1pt off 2pt,color=matlabyellow] (0,0) -- (8,4) -- (10,0);
\draw[color=matlabyellow] (8,4) node[above right] {$F_3$};

% F
\draw[thick] (0,0) -- (2,4) -- (5,5) -- (8,4) -- (10,0);
\draw (5,5) node[above] {$F=F_1 \supconv F_2 \supconv F_3$};

\end{tikzpicture}

\caption{Aggregate production function for three Leontief functions.}\label{fig:Leontief}
\end{figure}
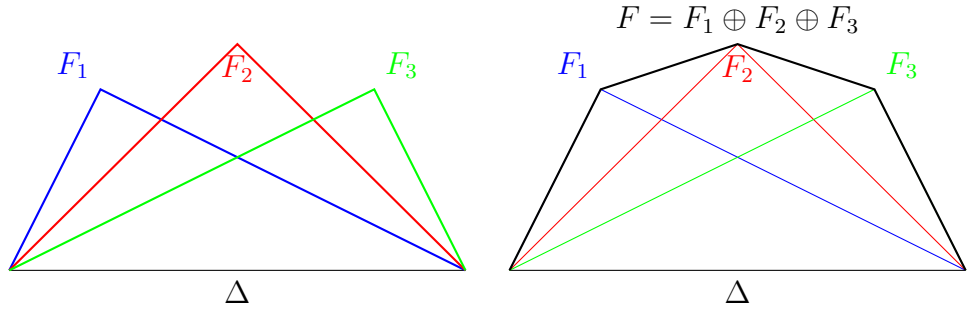
\end{exmp}

\subsection{Sparsity of optimal allocations}

In Figures \ref{fig:CD} and \ref{fig:Leontief}, every point on the graph of $F|_{\Delta_2}$ can be represented as a convex combination of at most two points drawn from the graphs of the $F_j|_{\Delta_2}$. This property is indeed general: the following proposition shows that, with $N$ inputs, the maximal output at any aggregate input can be attained using at most $N$ active production units.

\begin{prop}[Sparsity of optimal allocations]\label{prop:operate}
Suppose that each $F_j$ is homogeneous. For every $x\in \R_+^N$, there exists an allocation $(x_j)_{j=1}^J$ attaining $F(x)$ such that $\#\set{j:x_j\neq 0}\le N$.
\end{prop}

Thus, for every aggregate input vector $x$, the maximal output $F(x)$ can be attained by an allocation using at most $N$ active production units.

\section{Application to land and nonland factors}\label{sec:land}

The curvature of an aggregate production function can affect conclusions in growth and asset pricing models.  \citet{HiranoToda2025PNAS} study an economy in which land is both a factor of production and a store of value. Their Land Overvaluation Theorem shows that land overvaluation emerges when productivity grows sufficiently faster for nonland factors than for land and, among other conditions, the aggregate elasticity of substitution between land and nonland factors exceeds one at high input ratios.  This section shows how our linearity result provides a microfoundation for this technological condition.

Let $H$ denote a nonland factor, such as labor or human capital, and let $X$ denote land.  For a differentiable homogeneous production function $F:\R_{++}^2\to\R_+$ with positive marginal products, the elasticity of substitution can be written as
\begin{equation}
    \sigma_F(H,X)
    \coloneqq
    \left[-\frac{\partial\log(F_H/F_X)}{\partial\log(H/X)}\right]^{-1}. \label{eq:sigma_F}
\end{equation}
Homogeneity implies that the marginal rate of technological substitution $F_H/F_X$ depends only on the input ratio $H/X$, so the derivative is well defined.  If $F$ is linear with positive coefficients on a neighborhood of $(H,X)$, then $F_H/F_X$ is constant. The derivative in \eqref{eq:sigma_F} is then zero, and we interpret $\sigma_F=\infty$.

The following corollary identifies a simple condition under which a linearity cone contains all sufficiently high nonland-to-land input ratios.  The condition is that an efficient allocation jointly operates a land-using activity and a land-free activity.

\begin{cor}[Linearity at high nonland-to-land ratios]\label{cor:land}
Suppose $N=2$, write aggregate inputs as $(H,X)$, and suppose that the
hypotheses of Theorem \ref{thm:linear_profit} hold at a strictly positive
input price vector $w=(w_H,w_X)$.  Suppose an allocation attaining $F(x^*)$
contains a land-using activity and a land-free activity.  After relabeling
these as activities 1 and 2, respectively, write
\begin{equation*}
    x_1^*=(H_1,X_1),
    \qquad
    x_2^*=(H_2,0),
    \qquad
    X_1>0,\quad H_2>0.
\end{equation*}
Define $\kappa\coloneqq H_1/X_1$.  Then $F(H,X)=w_HH+w_XX$ whenever $H\ge \kappa X$. Consequently, $\sigma_F(H,X)=\infty$ for $X>0$ and $H>\kappa X$.
\end{cor}

\begin{proof}
For any $(H,X)$ satisfying $H\ge\kappa X$, noting that $H_1/X_1=\kappa$, we have
\begin{equation*}
    (H,X)
    =\frac{X}{X_1}(H_1,X_1)
    +\frac{H-\kappa X}{H_2}(H_2,0).
\end{equation*}
Both coefficients are nonnegative.  Therefore $(H,X)$ belongs to the cone generated by the activity-level input vectors in the allocation attaining $F(x^*)$.  Theorem \ref{thm:linear_profit} implies $F(H,X)=w_HH+w_XX$ on this cone.  At every point with $X>0$ and $H>\kappa X$, this equality holds on a neighborhood, so the inverse elasticity is zero and the elasticity is infinite.
\end{proof}

The presence of the land-free activity is important for the conclusion.  A generic full-dimensional cone generated by two strictly positive input vectors covers only a bounded interval of ratios $H/X$. Although the aggregate production function is linear on such a cone, the cone need not contain $(H,1)$ for large $H$.  By contrast, the vector $(H_2,0)$ lies on the nonland axis, so the cone in Corollary \ref{cor:land} contains every ratio above the threshold $\kappa$. Economically, once both activities operate, increases in the nonland factor beyond this threshold can be absorbed by expanding the land-free activity without changing the supporting factor prices.

Corollary \ref{cor:land} takes the coexistence of a land-using and a land-free activity as given.  In a two-sector economy, this coexistence can instead be derived from primitive properties of the land-using technology. The following proposition provides a sufficient condition.

\begin{prop}[Sufficient condition for high-ratio linearity]\label{prop:land_primitive}
Suppose there are two sectors.  Sector 1 has a concave, homogeneous, and increasing production function $F_1:\R_+^2\to\R_+$ such that $F_1(H,0)=0$, and sector 2 has the land-free technology $F_2(H,X)=aH$ for some $a>0$.  Define $f(h)\coloneqq F_1(h,1)$ and suppose that
\begin{equation*}
    b\coloneqq\max_{h\ge 0}\set{f(h)-ah}>0
\end{equation*}
is attained at some finite $h^*$.  Then, for $X>0$, the aggregate production function is
\begin{equation*}
    F(H,X)
    =aH+X\max_{0\le h\le H/X}\set{f(h)-ah}.
\end{equation*}
In particular, $F(H,X)=aH+bX$ whenever $H\ge h^*X$. In this region, an optimal allocation assigns $(h^*X,X)$ to sector 1 and $(H-h^*X,0)$ to sector 2; both sectors operate when $H>h^*X$. Consequently, $\sigma_F(H,X)=\infty$ for $X>0$ and $H>h^*X$.
\end{prop}

\begin{proof}
If $X=0$, the asserted linear formula follows immediately from $F_1(H,0)=0$.  Suppose henceforth that $X>0$. Because sector 2 does not use land and $F_1$ is nondecreasing, an optimal allocation can assign all land to sector 1.  If $H_1\in[0,H]$ is the nonland input assigned to that sector, homogeneity implies
\begin{align*}
    F(H,X)
    &=\max_{0\le H_1\le H}
      \set{F_1(H_1,X)+a(H-H_1)}\\
    &=aH+X\max_{0\le h\le H/X}\set{f(h)-ah},
\end{align*}
where $h=H_1/X$.  If $H\ge h^*X$, the global maximizer $h^*$ is feasible in the last maximization problem, so its value is $b$.  This proves the linear formula and shows that the stated allocation is optimal.  At every point with $X>0$ and $H>h^*X$, the formula holds on a neighborhood, and hence the elasticity of substitution is infinite.
\end{proof}

The condition in Proposition \ref{prop:land_primitive} has a simple marginal interpretation.  If $f$ is differentiable and strictly concave and $f'(0)>a>f'(\infty)$ holds, then $h^*$ is the unique finite solution to $f'(h^*)=a$.  The land-using sector employs $h^*X$ units of the nonland factor, and the land-free sector absorbs any amount in excess of $h^*X$.  Thus the aggregate marginal products remain constant above the threshold even when $F_1$ is strictly concave across input ratios.

\begin{exmp}[Cobb--Douglas and linear technologies]\label{ex:land_cobb_douglas}
We now apply Proposition \ref{prop:land_primitive} to the two-sector economy in \citet[\S4.1]{HiranoToda2025PNAS}.  Consider a land-intensive sector with Cobb--Douglas technology and a modern, land-free sector with a linear technology:
\begin{equation*}
    F_1(H,X)=A_1H^\alpha X^{1-\alpha},
    \qquad
    F_2(H,X)=A_2H,
\end{equation*}
where $A_1,A_2>0$ and $\alpha\in(0,1)$.  Define $\bar h\coloneqq (\alpha A_1/A_2)^{\frac{1}{1-\alpha}}$. Here $f(h)=A_1h^\alpha$ and $a=A_2$.  The strictly concave function $f(h)-ah$ has the unique maximizer $h^*=\bar h$, characterized by $\alpha A_1\bar h^{\alpha-1}=A_2$, and its maximal value is
\begin{equation*}
    b=(1-\alpha)A_1\bar h^\alpha>0.
\end{equation*}
Thus all the conditions of Proposition \ref{prop:land_primitive} hold. Because the second sector does not use land, all land can be assigned to the first sector.  If $H_1$ denotes the nonland input assigned to that sector, the aggregate production function is
\begin{equation*}
    F(H,X)
    =\max_{0\le H_1\le H}
    \set*{A_1H_1^\alpha X^{1-\alpha}+A_2(H-H_1)}.
\end{equation*}
The unconstrained maximizer is $H_1=\bar hX$.  It follows that
\begin{equation*}
    F(H,X)=
    \begin{cases*}
        A_1H^\alpha X^{1-\alpha}
        & if $H\le \bar hX$,\\
        A_2H+bX
        & if $H\ge \bar hX$.
    \end{cases*}
\end{equation*}
Below the threshold, only the land-intensive sector operates, and the aggregate elasticity of substitution is one.  Above the threshold, the land-intensive sector uses $(\bar hX,X)$ and the land-free sector uses $(H-\bar hX,0)$.  Thus, for this allocation, $\kappa=H_1/X_1=\bar h=h^*$: the thresholds in Corollary \ref{cor:land}, Proposition \ref{prop:land_primitive}, and this example coincide.  The aggregate production function is linear, with coefficient vector $w=(A_2,b)$, and its elasticity of substitution is infinite.  Thus aggregation changes the relevant elasticity from the unit elasticity of the land-using technology to infinite elasticity in the region where reallocation across sectors is possible.
\end{exmp}

In the structural transformation example of \citet[\S4.1]{HiranoToda2025PNAS}, the first sector represents a traditional land-intensive industry, such as agriculture, and the second represents a modern industry that uses labor or human capital but no land. Their sectoral productivities vary over time, so the threshold $\bar h$ also varies.  The calculation above shows that whenever both sectors operate, their aggregate technology is linear on a cone extending to arbitrarily high nonland-to-land ratios.  More generally, Corollary \ref{cor:land} shows that this conclusion does not depend on the Cobb--Douglas functional form: it follows from constant returns, profit maximization, and the coexistence of land-using and land-free activities.

This argument supplies a production-side foundation for the high-input elasticity condition in \citet{HiranoToda2025PNAS}; it does not by itself establish land overvaluation.  Their theorem also imposes conditions on relative productivity growth and its stochastic process.  The contribution here is to show that the required aggregate substitutability need not be a primitive property of a representative production unit.  It can arise endogenously through the allocation of inputs across heterogeneous activities.

\section{Conclusion}\label{sec:conclusion}

The aggregate production function gives the maximum total output attainable by allocating aggregate inputs across heterogeneous production units. Under constant returns to scale, any profit-maximizing allocation generates a cone on which the aggregate production function is linear.  Its dimension equals the dimension spanned by the units' input vectors in that allocation.  When the component technologies are concave, every strictly positive aggregate input admits such a cone, although it may be only a ray.  At a given supporting price, the maximal linearity cone is the sum of the component contact cones; it is full-dimensional exactly when the contact sets contain $N$ affinely independent input shares.

The geometric characterization complements this price-based result.  On the unit simplex, the aggregate production function is the least concave majorant of the pointwise maximum of the component technologies.  Nonconcavity of this pointwise maximum generates a higher-dimensional linearity cone.  Under strict concavity of the component technologies, the condition is also necessary, and with two inputs the cone is full-dimensional.  A dominant strictly concave technology provides the opposite case: aggregation reproduces that technology and yields no higher-dimensional linearity cone.  Moreover, maximal output can always be attained with at most as many active production units as there are inputs.  These results separate curvature inherited from the component technologies from curvature created by changes in the active set.

The application to land illustrates why the distinction matters.  When an efficient allocation combines a land-using activity with a land-free activity, the linearity cone contains all sufficiently high nonland-to-land input ratios. The aggregate elasticity of substitution is then infinite in that region, even when the land-using technology has finite elasticity.  Thus high aggregate substitutability can arise from reallocating inputs across production units rather than from substitution within a representative production unit.  With a concave land-using technology $F_1(H,X)=Xf(H/X)$ and a linear land-free technology $aH$, a finite maximizer of $f(h)-ah$ provides a primitive sufficient condition for this high-ratio linearity.  The exact linearity result relies on constant returns and frictionless input reallocation; capacity constraints, adjustment costs, or imperfect factor mobility may limit the relevant cones and restore aggregate curvature.

\appendix

\section{Proofs}\label{sec:proof}

\subsection{Proof of Lemma \ref{lem:agg_prod_homog}}
Take any $z\in \Delta_N$ and choose input vectors $y_j\in\R_+^N$, one for each firm, satisfying $\sum_{j=1}^J y_j=z$. Define $\alpha_j=\norm{y_j}\ge 0$ and $\alpha=(\alpha_1,\dots,\alpha_J)$. Then clearly $1=\norm{z}=\sum_{j=1}^J\norm{y_j}=\sum_{j=1}^J \alpha_j$, so $\alpha\in \Delta_J$. If $y_j\neq 0$, define $z_j=y_j/\norm{y_j}$. If $y_j=0$, define $z_j\in \Delta_N$ arbitrarily. Then clearly $z=\sum_{j=1}^J\alpha_jz_j$, and the choice of $z_j$ when $y_j=0$ is irrelevant. Conversely, every feasible choice $(\alpha_j,z_j)$ in \eqref{eq:agg_prod_homog} generates a feasible choice in \eqref{eq:agg_prod_func} by setting $y_j=\alpha_jz_j$. Since
\begin{equation*}
    \sum_{j=1}^J F_j(y_j)=\sum_{j=1}^J F_j(\alpha_jz_j)=\sum_{j=1}^J \alpha_jF_j(z_j)
\end{equation*}
by the homogeneity of $F_j$, the equivalence of \eqref{eq:agg_prod_func} and \eqref{eq:agg_prod_homog} holds. \hfill \qedsymbol

\subsection{Proof of Theorem \ref{thm:maximal_cone}}
\ref{item:maximal_cone1} Assigning all inputs to firm $j$ shows that
$F_j(x)\le F(x)\le w\cdot x$. We first show $Z(w)\supset Z_1(w)+\dotsb+Z_J(w)$. If $x_j\in Z_j(w)$ for every $j$, feasibility and the aggregate upper bound imply
\begin{equation*}
    F\left(\sum_{j=1}^Jx_j\right)
    \ge\sum_{j=1}^JF_j(x_j)
    =w\cdot\sum_{j=1}^Jx_j
    \ge F\left(\sum_{j=1}^Jx_j\right).
\end{equation*}
Therefore $\sum_jx_j\in Z(w)$. We next show $Z(w)\subset Z_1(w)+\dotsb+Z_J(w)$. Take $x\in Z(w)$ and an allocation $(x_j)$ attaining $F(x)$. Since
\begin{equation*}
    \sum_{j=1}^JF_j(x_j)=F(x)=w\cdot x
    =\sum_{j=1}^Jw\cdot x_j
\end{equation*}
and each term on the left is bounded above by the corresponding term on the right, $x_j\in Z_j(w)$ for every $j$.

\medskip
\noindent \ref{item:maximal_cone2} Under concavity, each $Z_j(w)$ is convex. Homogeneity makes $Z_j(w)$ a cone, and upper semicontinuity makes it closed. The same properties for $Z(w)$ follow directly from the concavity, homogeneity, and upper semicontinuity of $F$.

To show \eqref{eq:Zhatw}, take $x\in \widehat{Z}(w)$ and write $x=\sum_jx_j$ with $x_j\in Z_j(w)$. For $x_j\neq0$, set $\alpha_j=\norm{x_j}$ and $z_j=x_j/\alpha_j\in \widehat{Z}_j(w)$.  Then $\sum_j\alpha_j=1$ and $x=\sum_j\alpha_jz_j$, so $x$ belongs to the convex hull of $\bigcup_{j=1}^J\widehat{Z}_j(w)$.

Conversely, take
\begin{equation*}
    x\in\conv\left(\bigcup_{j=1}^J\widehat{Z}_j(w)\right).
\end{equation*}
Then there exist weights $\beta_k\ge0$ satisfying $\sum_{k=1}^K\beta_k=1$, indices $j(k)\in\set{1,\dots,J}$, and points $z_k\in\widehat{Z}_{j(k)}(w)$ such that $x=\sum_{k=1}^K\beta_kz_k$. For each firm $j$, define
\begin{equation*}
    x_j\coloneqq\sum_{\set{k:j(k)=j}}\beta_kz_k,
\end{equation*}
where an empty sum is zero. Because $\widehat{Z}_j(w)\subset Z_j(w)$ and $Z_j(w)$ is a convex cone, we have $x_j\in Z_j(w)$. Moreover, $x=\sum_{j=1}^Jx_j$. Part~\ref{item:maximal_cone1} therefore implies $x\in Z(w)$. Since $x$ is a
convex combination of points in $\Delta_N$, it also belongs to $\Delta_N$.
Hence $x\in Z(w)\cap\Delta_N=\widehat{Z}(w)$, which proves the reverse
inclusion.

\medskip
\noindent \ref{item:maximal_cone3}
The $\ell^1$ norm $\norm{\cdot}_1$ is linear on $\R_+^N$, so for a nonzero convex cone $C\subset \R_+^N$, the set $C\cap \Delta_N$ is a genuine hyperplane cross-section of affine dimension $\dim C-1$. Hence the dimension formula follows from homogeneous extension.  Finally, a convex subset of $\Delta_N$ has affine dimension $N-1$ if and only if it contains $N$ affinely independent points, proving the last claim. \hfill \qedsymbol

\subsection{Proof of Theorem \ref{thm:geo}}
Let $A=\hyp F|_{\Delta_N}$ and
$B=\conv\left(\bigcup_{j=1}^J\hyp F_j|_{\Delta_N}\right)$. To prove $A=B$, we prove $A\subset B$ and $B\subset A$.

Let $(x,y)\in A$, so $x\in \Delta_N$ and $y\le F(x)$. By Lemma
\ref{lem:agg_prod_homog}, choose $\alpha\in\Delta_J$ and normalized input
vectors $z_j\in\Delta_N$, one for each firm, such that
$x=\sum_{j=1}^J \alpha_jz_j$ and
$F(x)=\sum_{j=1}^J \alpha_jF_j(z_j)$. Therefore for each $j$, we can take
$y_j\le F_j(z_j)$ such that $y=\sum_{j=1}^J \alpha_jy_j$. By the definition
of the hypograph, we have $(z_j,y_j)\in \hyp F_j|_{\Delta_N}$ and
$(x,y)=\sum_{j=1}^J \alpha_j(z_j,y_j)$, so
\begin{equation*}
    (x,y)\in \conv\left(\bigcup_{j=1}^J\hyp F_j|_{\Delta_N}\right)=B.
\end{equation*}
Therefore $A\subset B$.

Conversely, let $(x,y)\in B$. By the definition of the convex hull, we can take $\beta\in \Delta_K$, and for each $k\in\set{1,\dots,K}$, indices $j(k)\in \set{1,\dots,J}$ and points $(z_k,t_k)\in \hyp F_{j(k)}|_{\Delta_N}$ such that $(x,y)=\sum_{k=1}^K \beta_k(z_k,t_k)$. In particular,
\begin{align}
    x&=\sum_{k=1}^K\beta_k z_k, & y&=\sum_{k=1}^K\beta_k t_k, & t_k&\le F_{j(k)}(z_k). \label{eq:xyt}
\end{align}
Since each $z_k\in\Delta_N$ and $\Delta_N$ is convex, we have
$x\in\Delta_N$.

For each firm $j$, define $x_j\coloneqq
\sum_{\set{k:j(k)=j}}\beta_k z_k$, where an empty sum is understood to be zero. Then $x_j\in\R_+^N$ and $\sum_{j=1}^Jx_j=\sum_{k=1}^K\beta_kz_k=x$. Thus the vectors $x_j$, one for each firm, form a feasible allocation of the aggregate input
$x$.

Because $F_j$ is concave and homogeneous, it is superadditive (see \eqref{eq:superadditive}). Consequently, homogeneity, superadditivity, and \eqref{eq:xyt} imply
\begin{align*}
    F_j(x_j)&=F_j\left(\sum_{\set{k:\,j(k)=j}}\beta_kz_k\right) \ge \sum_{\set{k:\,j(k)=j}}F_j(\beta_kz_k) \\
    &=\sum_{\set{k:\,j(k)=j}}\beta_kF_j(z_k) \ge \sum_{\set{k:\,j(k)=j}}\beta_kt_k.
\end{align*}
Summing over $j$ and using the definition of the aggregate production
function and \eqref{eq:xyt}, we obtain
\begin{equation*}
    F(x) \ge \sum_{j=1}^JF_j(x_j) \ge \sum_{k=1}^K\beta_kt_k=y.
\end{equation*}
Therefore $(x,y)\in\hyp F|_{\Delta_N}=A$ and $B\subset A$. \hfill \qedsymbol

\subsection{Proof of Theorem \ref{thm:M}}
\ref{item:M1}
Suppose $M$ is not concave.  Since $F|_{\Delta_N}$ is its least concave majorant by Theorem \ref{thm:geo}, there exists $z\in\Delta_N$ such that $F(z)>M(z)$.  By Lemma \ref{lem:agg_prod_homog}, after omitting zero coefficients, we may write
\begin{equation}
    z=\sum_{j\in A}\alpha_jz_j,
    \qquad
    F(z)=\sum_{j\in A}\alpha_jF_j(z_j), \label{eq:Fz_decompose}
\end{equation}
where $\alpha_j>0$ for $j\in A$, $\sum_{j\in A}\alpha_j=1$, and $z_j\in\Delta_N$.  Because $F_j(z_j)\le M(z_j)\le F(z_j)$ and $F$ is concave, we have
\begin{equation}
    F(z)=\sum_{j\in A}\alpha_jF_j(z_j)
    \le\sum_{j\in A}\alpha_jF(z_j)
    \le F(z). \label{eq:Fz_chain}
\end{equation}
All inequalities are therefore equalities.  Equality in the concavity inequality implies that $F$ is affine on $K\coloneqq\conv\set{z_j:j\in A}$.  To verify this implication, take any $y\in K$ and choose coefficients $\beta_j\ge0$ such that $\sum_{j\in A}\beta_j=1$ and $y=\sum_{j\in A}\beta_jz_j$.  Since every $\alpha_j$ is strictly positive, we can choose $\epsilon\in(0,1)$ small enough that $\epsilon\beta_j\le\alpha_j$ for every $j\in A$.  Define
\begin{equation*}
    \gamma_j\coloneqq
    \frac{\alpha_j-\epsilon\beta_j}{1-\epsilon}\ge 0,
    \qquad
    y'\coloneqq\sum_{j\in A}\gamma_jz_j.
\end{equation*}
Then $\sum_{j\in A}\gamma_j=1$ and
$z=\epsilon y+(1-\epsilon)y'$. Therefore,
\begin{align*}
    F(z)
    &\ge \epsilon F(y)+(1-\epsilon)F(y') && (\because \text{$F$ is concave}) \\
    &\ge \epsilon\sum_{j\in A}\beta_jF(z_j)
       +(1-\epsilon)\sum_{j\in A}\gamma_jF(z_j) && (\because \text{$F$ is concave})\\
    &=\sum_{j\in A}\alpha_jF(z_j)=F(z). && (\because \eqref{eq:Fz_chain})
\end{align*}
Thus equality holds throughout, and in particular $F(y)=\sum_{j\in A}\beta_jF(z_j)$.  Since $y$ was arbitrary, $F$ is affine on $K$. At least two of the points $z_j$ are distinct; otherwise \eqref{eq:agg_prod_homog} and \eqref{eq:Mz} would give $M(z)=F(z)$, contradicting the assumption $F(z)>M(z)$. The segment joining any two distinct such points is therefore a nondegenerate segment on which $F$ is affine. Because the segment lies in $\Delta_N$, this affine function can be written as $a\cdot z$ for some $a\in\R^N$. Homogeneity therefore gives $F(x)=a\cdot x$ on the convex cone generated by the segment.  Its endpoints are distinct points on $\Delta_N$ and hence are linearly independent, so this is a higher-dimensional linearity cone.

\medskip
\noindent \ref{item:M2}
It suffices to show that if each $F_j|_{\Delta_N}$ is strictly concave and $M$ is concave, then $F$ has no higher-dimensional cone. Since $M$ is concave,  $F|_{\Delta_N}=M$. Suppose to the contrary that $F$ had a higher-dimensional linearity cone $C$. The convex set $C\cap\Delta_N$ then has affine dimension $\dim C-1\ge1$, so it contains a nondegenerate line segment on which $F|_{\Delta_N}=M$ is affine.  On this segment, every $F_j$ lies below the affine function $M$. Strict concavity implies that a given $F_j$ can touch this affine upper bound at no more than one point: equality at two distinct points would make $F_j$ strictly exceed the bound between them. The finite collection of component technologies therefore cannot attain $M$ at every point of the segment, a contradiction. Hence no higher-dimensional linearity cone can exist. \hfill \qedsymbol

\subsection{Proof of Proposition \ref{prop:operate}}
The proof uses a technique similar to that used in the proof of the Carath\'eodory theorem. The claim is immediate if $x=0$. Suppose that $x\neq 0$, and set $z\coloneqq x/\norm{x}\in\Delta_N$. Lemma \ref{lem:agg_prod_homog} yields
$\alpha\in\Delta_J$ and $z_j\in\Delta_N$ satisfying \eqref{eq:Fz_decompose} with an index set $A$, where $\alpha_j>0$ for $j\in A$. Suppose that $\#A>N$. Since the vectors $(z_j)_{j\in A}$ belong to $\R^N$, they are linearly dependent. Hence there exists a nonzero vector $d=(d_j)\in \R^J$ with $d_j=0$ if $j\notin A$ such that $\sum_{j=1}^Jd_jz_j=0$. Because $z_j\in\Delta_N$, taking the sum of the coordinates gives $\sum_{j=1}^Jd_j=0$. Thus $d$ has both positive and negative components.

Define $c\coloneqq \sum_{j=1}^Jd_jF_j(z_j)$. By replacing $d$ by $-d$ if necessary, we may assume $c\ge 0$. Define the affine function
\begin{equation*}
    \phi(t)\coloneqq \sum_{j=1}^J(\alpha_j+td_j)F_j(z_j)=F(z)+ct
\end{equation*}
and set $t^*=\min_{\set{j:d_j<0}}(-\alpha_j/d_j)>0$. By the properties of $d$ and $t^*$, we have $\alpha+t^*d\in \Delta_J$ and $\sum_{j=1}^J(\alpha_j+t^*d_j)z_j=z$. Since $t^*>0$ and $c\ge 0$, we have
\begin{equation*}
    \phi(t^*)\ge\phi(0)=F(z).
\end{equation*}
Since $\phi(t^*)$ is attainable at the aggregate input $z$, it cannot exceed $F(z)$. Hence $\phi(t^*)=F(z)$. Since $\alpha_j+t^*d_j=0$ for some $j\in A$, at least one previously positive coefficient becomes zero at
$t^*$.

Repeating this argument yields an optimal representation of
$z$ with at most $N$ positive coefficients. With a slight abuse of notation, let $(\alpha_j)_{j=1}^J$ denote the resulting coefficients
and define $x_j\coloneqq\norm{x}\alpha_jz_j$. Then $\sum_jx_j=x$, at most $N$ of the vectors $x_j$ are nonzero,
and homogeneity gives
\begin{equation*}
    \sum_{j=1}^JF_j(x_j)=\norm{x}\sum_{j=1}^J\alpha_jF_j(z_j)=\norm{x}F(z)=F(x). \tag*{\qedsymbol}
\end{equation*}

\printbibliography

\end{document}